\documentclass[prx,a4paper,aps,twocolumn,superscriptaddress,10pt,bibnotes]{revtex4-2}

\usepackage{graphicx,color}
\usepackage{dcolumn}
\usepackage{bm}
\usepackage{amsmath,amssymb,mathrsfs}
\usepackage{url}
\usepackage{hyperref}
\usepackage{caption}
\usepackage{overpic}
\usepackage{subcaption}
\usepackage{ragged2e} 

\def\>{\rangle}
\def\<{\langle}

\newcommand{\Tr}{\operatorname{Tr}}

\begin{document}
\title{Experimental observation of entropic-singularity-induced nonadditive quantum communication in a qutrit platypus channel}
    
\author{Yu Guo}
\thanks{These two authors contributed equally to this work.}
\affiliation{Laboratory of Quantum Information, University of Science and Technology of China, Hefei 230026, China}
\affiliation{CAS Center for Excellence in Quantum Information and Quantum Physics, 	University of Science and Technology of China, Hefei 230026, China}
\affiliation{Anhui Province Key Laboratory of Quantum Network, University of Science and Technology of China, Hefei 230026, China}

\author{Bo-Xuan Wang}
\thanks{These two authors contributed equally to this work.}
\affiliation{Laboratory of Quantum Information, University of Science and Technology of China, Hefei 230026, China}
\affiliation{CAS Center for Excellence in Quantum Information and Quantum Physics, 	University of Science and Technology of China, Hefei 230026, China}
\affiliation{Anhui Province Key Laboratory of Quantum Network, University of Science and Technology of China, Hefei 230026, China}
    
\author{Xiao-Min Hu}
\affiliation{Laboratory of Quantum Information, University of Science and Technology of China, Hefei 230026, China}
\affiliation{CAS Center for Excellence in Quantum Information and Quantum Physics, 	University of Science and Technology of China, Hefei 230026, China}
\affiliation{Anhui Province Key Laboratory of Quantum Network, University of Science and Technology of China, Hefei 230026, China}
\affiliation{Hefei National Laboratory, University of Science and Technology of China, Hefei 230088, China}

\author{Yun-Feng Huang}
\affiliation{Laboratory of Quantum Information, University of Science and Technology of China, Hefei 230026, China}
\affiliation{CAS Center for Excellence in Quantum Information and Quantum Physics, 	University of Science and Technology of China, Hefei 230026, China}
\affiliation{Anhui Province Key Laboratory of Quantum Network, University of Science and Technology of China, Hefei 230026, China}
\affiliation{Hefei National Laboratory, University of Science and Technology of China, Hefei 230088, China}
    
\author{Chuan-Feng Li}
\affiliation{Laboratory of Quantum Information, University of Science and Technology of China, Hefei 230026, China}
\affiliation{CAS Center for Excellence in Quantum Information and Quantum Physics, 	University of Science and Technology of China, Hefei 230026, China}
\affiliation{Anhui Province Key Laboratory of Quantum Network, University of Science and Technology of China, Hefei 230026, China}
\affiliation{Hefei National Laboratory, University of Science and Technology of China, Hefei 230088, China}

\author{Guang-Can Guo}
\affiliation{Laboratory of Quantum Information, University of Science and Technology of China, Hefei 230026, China}
\affiliation{CAS Center for Excellence in Quantum Information and Quantum Physics, 	University of Science and Technology of China, Hefei 230026, China}
\affiliation{Anhui Province Key Laboratory of Quantum Network, University of Science and Technology of China, Hefei 230026, China}
\affiliation{Hefei National Laboratory, University of Science and Technology of China, Hefei 230088, China}

\author{Bi-Heng Liu}
\email{bhliu@ustc.edu.cn}
\affiliation{Laboratory of Quantum Information, University of Science and Technology of China, Hefei 230026, China}
\affiliation{CAS Center for Excellence in Quantum Information and Quantum Physics, 	University of Science and Technology of China, Hefei 230026, China}
\affiliation{Anhui Province Key Laboratory of Quantum Network, University of Science and Technology of China, Hefei 230026, China}
\affiliation{Hefei National Laboratory, University of Science and Technology of China, Hefei 230088, China}
\affiliation{College of Physics, Guizhou University, Guiyang 550025, China}

\begin{abstract}
The nonadditivity of channel capacity is a defining feature that distinguishes quantum communication from classical communication. In the quantum realm, the channel capacity is determined by coherent information, which is defined through the von Neumann entropies of the output and its environment. Despite its fundamental importance, experimental evidence of such nonadditive quantum communication has been elusive because of the complexity of the required quantum channel. Here, we experimentally observe entropic-singularity-induced coherent-information nonadditivity using the qutrit platypus channel implemented on a photonic platform. By preparing six-dimensional photonic entanglement, we directly measure the coherent information of a platypus channel, a qubit amplitude damping channel, and their joint uses, revealing a clear violation of additivity. Quantum process tomography further reveals the entropic singularity responsible for this effect, demonstrating how singular entropy landscapes in low-dimensional channels can enhance quantum communication beyond additive limits.
\end{abstract}

\maketitle

\textit{Introduction.---} Communication theory fundamentally addresses the reliable transmission of information through noisy channels. In the classical regime, the ultimate rate of error-free transmission is characterized by the Shannon capacity~\cite{shannon1948mathematical}, defined through the maximum mutual information between channel inputs and outputs and governed by the Shannon entropy. Quantum communication extends this framework to quantum systems, enabling the transmission of classical~\cite{holevo1973bounds,schumacher1997sending,holevo2002capacity}, entanglement-assisted classical~\cite{bennett1996mixed,bennett1999entanglement}, private~\cite{cai2004quantum,devetak2005private}, and quantum information~\cite{bennett1996mixed,devetak2005private,lloyd1997capacity,divincenzo1998quantum,barnum2002quantum}. The ultimate rate for faithfully transmitting quantum information—the quantum capacity~\cite{bennett1996mixed,devetak2005private,lloyd1997capacity,divincenzo1998quantum,barnum2002quantum,macchiavello2016detecting,leditzky2018dephrasure}—is determined by the coherent information~\cite{schumacher1996quantum}, involving the von Neumann entropies of the channel output and its complementary environment. This transition from the Shannon to the von Neumann entropy is a fundamental departure from classical information theory.

A striking consequence of this departure is the nonadditivity of channel capacities~\cite{shor9604006quantum,divincenzo1998quantum,leditzky2018dephrasure,smith2007degenerate,smith2009extensive,smith2009can,elkouss2015superadditivity,hastings2009superadditivity,smith2008quantum,oppenheim2008quantum,smith2011quantum,brandao2012does,lim2019activation,horodecki2005secure,leung2014maximal}. In classical information theory, capacities are strictly additive: two channels used in parallel support a total transmission rate equal to the sum of their individual capacities. Quantum channels, however, can violate this rule such that joint channel use enables higher communication rates than independent usage does. This breakdown of additivity reflects the collective nature of quantum information transmission, arising from the nonlinear entropy dependence of coherent information and entangled channel encoding. Understanding the physical origin of this nonadditivity is central to identifying the ultimate limits of quantum communication.

However, nonadditivity remains difficult to characterize and observe experimentally. Coherent information is neither convex nor additive, and quantum capacity generally requires regularization over arbitrarily many channel uses, rendering analytical evaluation intractable for most channels. Experimentally, coherent information depends on both the channel output and its complementary environment, which are difficult to access simultaneously, whereas nonadditive effects are often subtle and obscured by loss, decoherence, and finite statistics. As a result, experimental demonstrations of nonadditivity have so far been limited to carefully engineered channel models~\cite{leditzky2018dephrasure,yu2020experimental}.

A central open question concerns the role of entropic singularities. In several theoretical constructions that exhibit nonadditivity, the coherent information displays abrupt variations associated with the logarithmic singular behavior of the von Neumann entropy~\cite{siddhu2021entropic}. Such singular responses reflect the extreme sensitivity of the von Neumann entropy to infinitesimal state perturbations, a feature with no classical analogue. These entropic singularities have been shown to give rise to both positive coherent information and superadditivity, most transparently in the so-called platypus channel~\cite{siddhu2021entropic,leditzky2023generic,leditzky2023platypus}. Introduced as a low-dimensional quantum channel interpolating between degradable and antidegradable regimes, this channel constitutes the simplest known example in which coherent information is governed by logarithmic entropic singularities. Importantly, its structure admits a clear separation between the output and environmental entropy responses, making it an ideal platform for probing the physical origin of nonadditivity.

In this work, we experimentally demonstrate entropic-singularity-induced nonadditivity in quantum communication using a $3$-dimensional platypus channel combined with a qubit amplitude damping channel. We implement the channels on a photonic platform with process fidelities exceeding $0.98$ and probe their entropic structure using quantum process tomography. By preparing $6$-dimensional photonic entanglement as input, we perform a full spectral reconstruction of the output states, revealing that the logarithmic entropy singularity of the channel output dominates over that of its complementary environment. From the reconstructed density matrices, we directly evaluate the von Neumann entropies and coherent information and observe a violation of additivity by more than $20$ standard deviations, where the joint coherent information exceeds the sum of the individual contributions. These results provide direct experimental identification of the entropic mechanism underlying nonadditivity and establish a scalable approach for accessing collective quantum communication advantages in low-dimensional systems.

\textit{Background: Quantum capacity and nonadditivity.---} In a standard communication scenario, information is encoded into the internal state of a quantum particle and transmitted to a receiver through a noisy channel. The ability of a channel to faithfully transmit quantum information is quantified by its quantum capacity, defined as the maximum number of qubits that can be reliably transmitted per channel use. In quantum information theory, the quantum capacity is lower bounded by single-use coherent information, which is defined as follows:
\begin{equation}\label{coheinformation}
    Q^{(1)}(\mathcal{B})=\max_{\rho}[S\left(\mathcal{B}\left(\rho\right)\right)-S(\mathcal{B}^{c}\left(\rho\right))],
\end{equation}
where $\rho$ is the input state, $\mathcal{B}^{c}$ is the complementary channel, and $S(\rho)=-\Tr{(\rho\log\rho})$ denotes the von Neumann entropy on $\rho$. The quantum capacity is obtained from regularized coherent information $Q(\mathcal{B})=\lim_{n\to\infty}\frac{1}{n}Q^{(1)}(\mathcal{B}^{\otimes n})$, reflecting the possibility of collective encoding and decoding across multiple channel uses. 

A key distinction between classical and quantum communication lies in the potential nonadditivity of coherent information and quantum capacity. For two quantum channels $\mathcal{B}_{1}$ and $\mathcal{B}_{2}$, the superadditivity of coherent information occurs when
\begin{equation}\label{nonadditivity}
    Q^{(1)}(\mathcal{B}_1\otimes\mathcal{B}_2)>Q^{(1)}(\mathcal{B}_1)+
Q^{(1)}(\mathcal{B}_2),
\end{equation}
with a similar relation defining the superadditivity of the quantum capacity. This effect signals a breakdown of strong additivity, demonstrating that the joint use of distinct quantum channels can transmit quantum information more efficiently than independent usage can.

\textit{Entropic singularity and platypus channel.---} The coherent information, thus the quantum capacity, of a quantum channel is ultimately rooted in the behavior of von Neumann entropy. Unlike the Shannon entropy, the von Neumann entropy is generally nonsmooth because small variations in a quantum state can induce large changes in entropy. This nonsmoothness becomes particularly pronounced when the eigenvalues of the state approach zero, giving rise to so-called \emph{$\log$-singularity}~\cite{siddhu2021entropic}. 

More precisely, consider a family of quantum states $\rho(\epsilon)$ parameterized by a parameter $\epsilon$, such that one or more eigenvalues of $\rho(\epsilon)$ increase linearly from zero to the leading order in $\epsilon$. In this regime, it has been proven that the entropy exhibits a logarithmic divergence in its derivative,
\begin{equation}\label{singularity}
    \frac{dS(\epsilon)}{d\epsilon} \sim -x \log \epsilon,
\end{equation}
where the coefficient $x$ characterizes the strength of the singularity. Physically, $x$ quantifies the sensitivity of the entropy response to small quantum state perturbations.

The relevance of entropic $\log$-singularity to quantum information transmission becomes apparent when coherent information (\ref{coheinformation}) is considered. Because both terms can exhibit singular behavior, the magnitudes of coherent information are governed by the imbalance between the corresponding singularity rates. When the entropy of channel $\mathcal{B}$ has a stronger $\log$-singularity rate than that of its complementary channel $\mathcal{B}^c$ does, the coherent information $Q^{(1)}(\mathcal{B})$ can become positive. Such imbalance provides a physical mechanism for the emergence of a positive quantum capacity. The same reasoning extends to nonadditivity effects where channels are used jointly. When two channels are used in parallel, the combined output and environment entropies may exhibit $\log$-singularities that are absent or weaker in the individual coherent information, leading to superadditivity.

A paradigmatic realization of this mechanism is provided by the $d$-dimensional platypus channel $M_{d}$. This channel is defined by the superoperator $M_{d}(\rho)=\Tr_{c}(G\rho G^\dagger)$, where the isometry $G: a\mapsto b\otimes c$ takes the form
\begin{align}
G|0\rangle &=\frac{1}{\sqrt{d-1}}\sum_{i=0}^{d-2}|i\rangle\otimes|i\rangle \nonumber\\
G|j\rangle &=|d-1\rangle\otimes|j-1\rangle, \, for \, j=1,2,\dots,d-1. \label{platypus}
\end{align}
The channel $M_{d}$ acts on a $d$-dimensional input and output Hilbert space, while its environment has dimension $d-1$, with $d\geq 3$. From the entropic perspective, $M_{d}$ features a $\log$-singularity in its output entropy whose rate exceeds that of its complementary channel, thereby enabling positive coherent information. More importantly, when used jointly with some well-known qubit channels, such as depolarizing channel, erasure channel, and amplitude damping channel, the combined channel displays a superadditivity of coherent information or even quantum capacity.

\begin{figure*}[htbp]
    \centering
    \includegraphics[width=1.5\columnwidth]{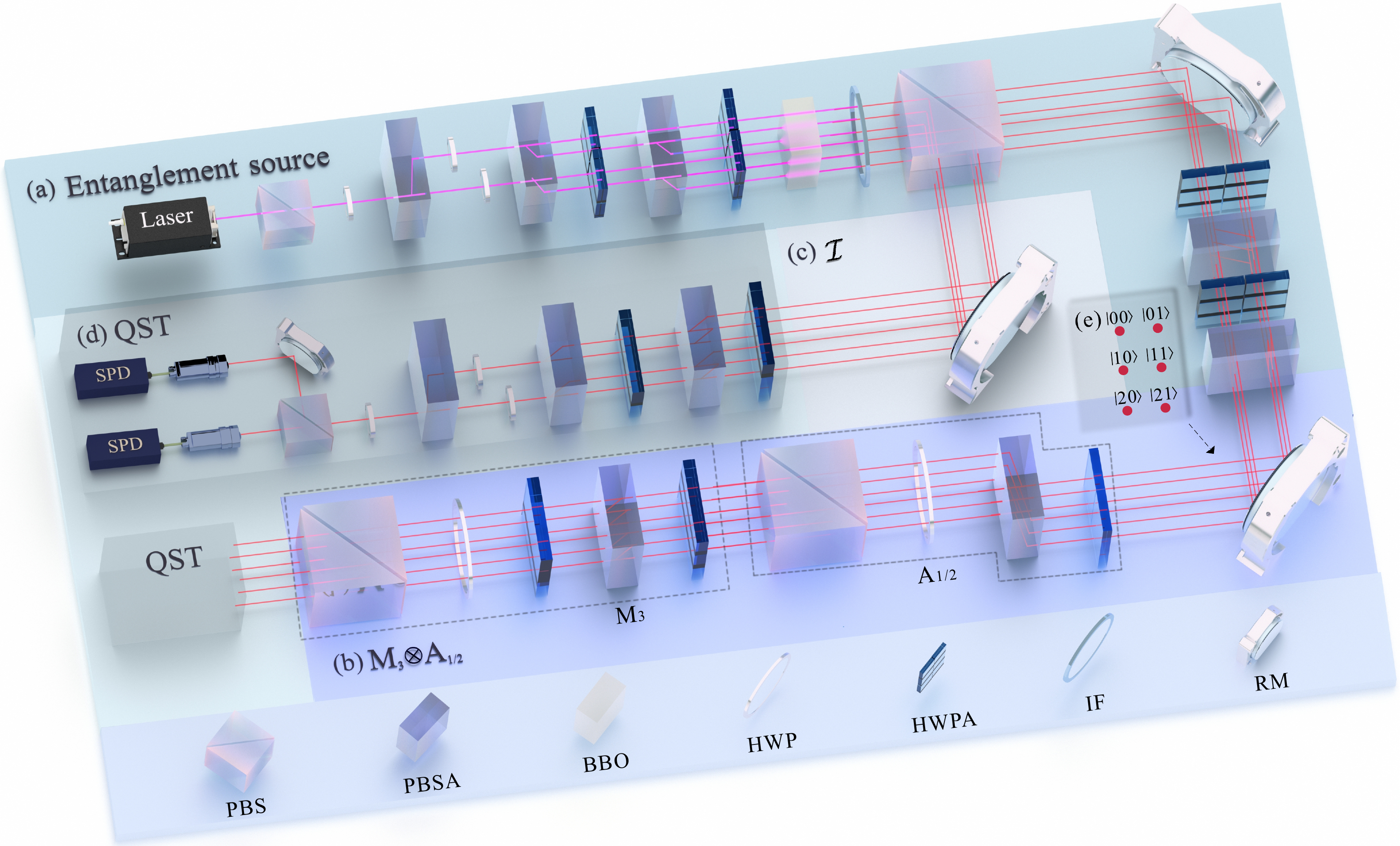}
        \caption{\justifying{\em Experimental setup.}  (a) Preparation of the 6-dimensional entangled state. A 404 nm pump laser is expanded into a $2\times3$ parallel beam array, which coherently pumps a BBO crystal to generate entangled photon pairs via SPDC. The two photons are separated by a PBS, after which one photon undergoes a local unitary operation to achieve the target input states for the channels. (b) Realization of the channel $M_3\otimes A_{1/2}$ using PBSAs, HWPAs, and a PBS. (d) Quantum state tomography of the 6-dimensional quantum state. (e) Encoding strategy of the six-dimensional path modes. PBS, polarizing beam splitter; HWP, half-wave plate; IF, interference filter; RM, reflection mirror; SPD, single-photon detector.
        }
      \label{Expsetup}
\end{figure*}

\textit{Experimental Setup.---} We present an experimental demonstration of the entropic-singularity-induced nonadditive quantum communication using a $3$-dimensional platypus channel $M_{3}$ combined with a qubit amplitude damping channel, $A_\gamma (\rho)=K_0\rho K_0^\dagger+K_1\rho K_1^\dagger$ with a damping parameter $\gamma=\frac{1}{2}$ and Kraus operators $K_0=|0\>\<0|+\sqrt{1-\gamma}|1\>\<1|$ and $K_1=\sqrt{\gamma}|0\>\<1|$. Our experimental setup, given in Fig.~\ref{Expsetup}, features (a) a six-dimensional photonic entanglement source, (b) a product channel $M_3\otimes A_{1/2}$, (c) an identical channel, and (d) quantum state tomography modules. 

The core component of our setup is the implementation of the joint channel $M_3\otimes A_{1/2}$. Our encoding scheme, illustrated in the inset of Fig.~\ref{Expsetup}, employs a $3\times 2$ spatial mode array: the three rows encode the input states $|0\>$, $|1\>$, and $|2\>$ of the $M_3$ channel, while the two columns encode the input states $|0\>$ and $|1\>$ of the $A_{1/2}$ channel. 
The realizations of the $M_3$ and $A_{1/2}$ channels are implemented via the assemblages indicated by the two boxes in Fig.~\ref{Expsetup}, which act on the rows and columns of the mode array, respectively. Taking $M_3$ as an example, its action can be described by the following Kraus operators: $M_3(\rho)=K_0\rho K_0^\dagger+K_1\rho K_1^\dagger$, with $K_0=|0\>\<0|/\sqrt{2}+|2\>\<1|$ and $K_1=|1\>\<0|/\sqrt{2}+|2\>\<2|$. The overall effect is that the input state $|0\>$ is mapped to a mixed state of $|0\>$ and $|1\>$, while the input states $|1\>$ and $|2\>$ are both mapped to $|2\>$. In the $M_3$ assemblage, the first half-wave plate array (HWPA) and polarizing beam splitter array (PBSA) control the relative amplitudes of the beams in each row to satisfy the above input-output relations, whereas the second HWPA adjusts the polarization of the beams so that the PBS transmission and reflection ports correspond to the two Kraus operators acting on the input states. The implementation of the $A_{1/2}$ channel follows a similar principle. Further experimental details are provided in the SM~\cite{SM}.  Notably, our method for implementing the Platypus channel can be readily generalized to higher-dimensional cases (see SM~\cite{SM}).

\begin{figure*}[htbp]
    \centering
    \begin{minipage}[t]{0.46\textwidth}
        \centering
        \begin{overpic}[width=\linewidth]{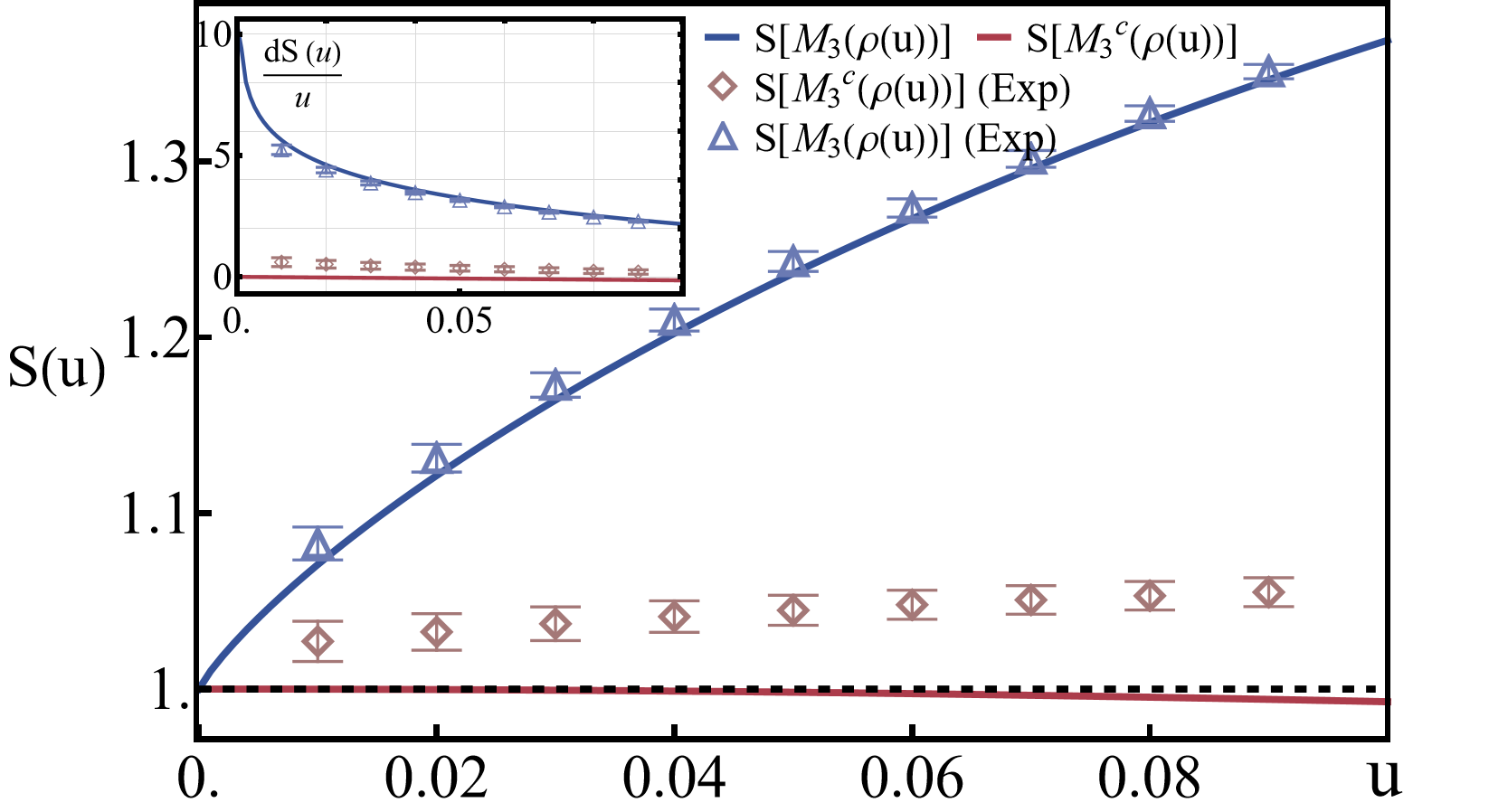}
            \put(50,-2){\bfseries (a)}
        \end{overpic}
    \end{minipage}
    \hspace{0.01\textwidth}
    \begin{minipage}[t]{0.46\textwidth}
        \centering
        \begin{overpic}[width=\linewidth]{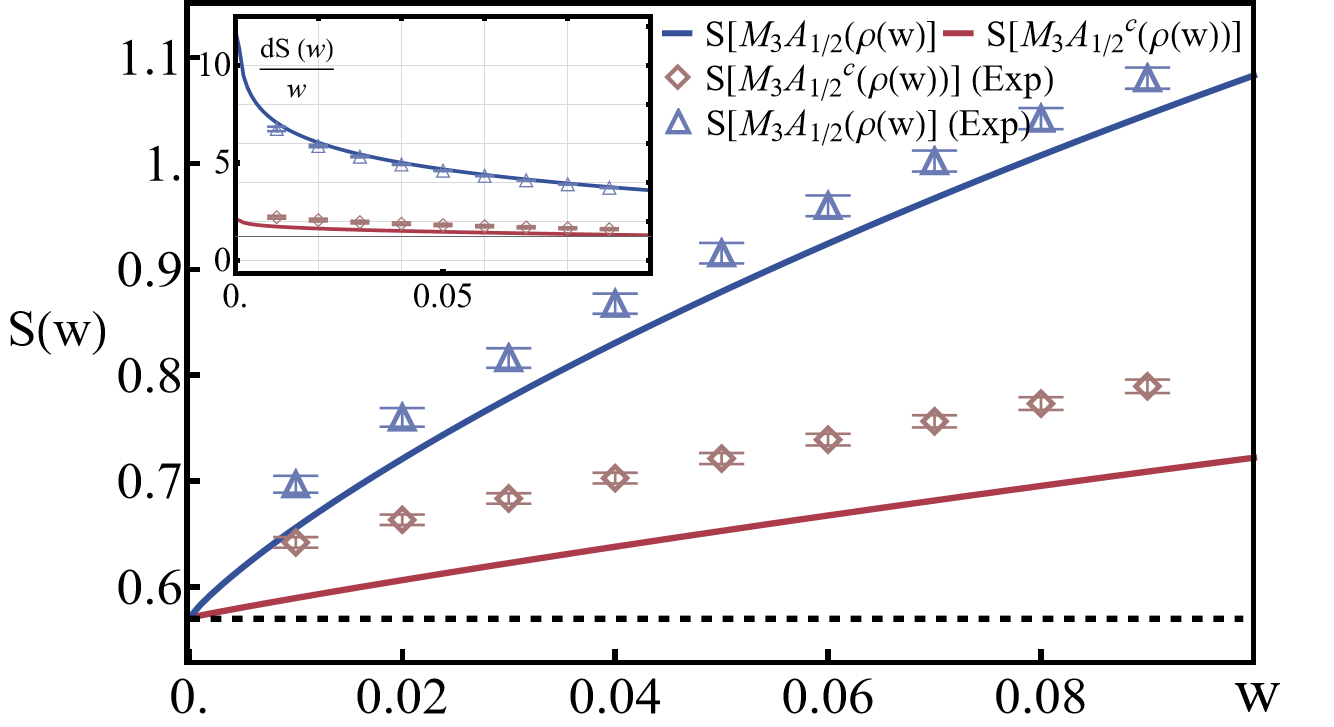}
            \put(50,-2){\bfseries (b)}
        \end{overpic}
    \end{minipage}

    \caption{\justifying\emph{Experimental results on the entropy singularity of (a) $\mathcal{M}_{3}$ and (b) $\mathcal{M}_{3}\otimes\mathcal{A}_{1/2}$.} For the input states $\rho(u)$ and $\rho(w,v)$ with the parameters $u$ and $w$ varied, respectively, the measured von Neumann entropies of the channel outputs and their complementary channels are marked as triangles and diamonds, while the corresponding theoretical predictions are plotted as solid blue and orange lines, respectively. The insets display the quantities $\frac{dS(u)}{u}$ and $\frac{dS(w)}{w}$ in Eq.~(\ref{singularity}) for $u\in[0, 0.1]$ and $w\in[0, 0.1]$, respectively.}
    \label{fig:singularity}
\end{figure*}

To measure the coherent information of the channel, we optimize the input state $\rho$ and reconstruct the output density matrix via quantum state tomography, from which the corresponding von Neumann entropy $S\left(M_3\otimes A_{1/2}\left(\rho\right)\right)$ is evaluated. To determine the complementary term $S\left(M_3\otimes A_{1/2}^{c}\left(\rho\right)\right)$, we purify the input state as $\rho'$, with $\rho=Tr(\rho')$. The measurement of the complementary contribution is therefore reduced to measuring $S(M_3\otimes A_{1/2}\otimes \mathcal{I}(\rho'))$, in which the identical channel $\mathcal{I}$ acts as the complementary channel. To this end, we implement a six-dimensional entangled state using the setup shown in Fig.~\ref{Expsetup}. The central idea is to employ HWPs and PBSAs to expand the number of path modes of a 404 nm pump laser, which then coherently pumps a BBO crystal to generate entangled photon pairs via spontaneous parametric downconversion (SPDC). The two photons are subsequently separated by a PBS, after which one photon undergoes a local unitary transformation to prepare the target entangled state. The two photons are then sent through the channel $M_3\otimes A_{1/2}$ and its complementary channel $\mathcal{I}$, respectively, followed by quantum state tomography measurements. Further experimental details are provided in the SM~\cite{SM}.

\textit{Results analysis.---} As the first result, we characterize the $\log$-singularity properties of $M_3$ and $M_3\otimes A_{1/2}$. We perform quantum process tomography to reconstruct both channels with fidelities of $0.997\pm0.003$ and $0.983\pm 0.003$ (see SM for details). For channel $M_3$ ($M_3\otimes A_{1/2}$) and its complementary channel $M_3^c$ ($(M_3\otimes A_{1/2})^c$), we parameterize the input states as $\rho(u)=(1-u)[0]+u[2]$ ($\rho(w,v)=(1-w)[00]+w[\xi]$) with $u\in[0,0.1]$ ($w\in[0,0.1]$, $|\xi\>=\sqrt{1-v}|20\>+\sqrt{v}|11\>$), where $[0]$ denotes $|0\rangle\langle0|$, and compute the eigenvalues and von Neumann entropies of the corresponding output states. The experimentally measured entropies of $M_3$ ($M_3\otimes A_{1/2}$) and $M_3^c$ ($(M_3\otimes A_{1/2})^c$), denoted as $S(u)$($S(w)$) for simplicity, are shown as triangles and diamonds in Fig.~\ref{fig:singularity}, while the theoretical predictions are given by the blue and red solid lines. The difference between the output entropy of a channel and that of its complementary channel yields coherent information, from which we observe positive coherent information for both channels on these families of input states.

To further substantiate this observation, we quantitatively extract the $\log$-singularity strength of $M_3$ and $M_3\otimes A_{1/2}$, as shown in the inset of Fig.~\ref{fig:singularity}. The extracted $\log$-singularity strengths of the output channels are $0.998\pm0.005$ and $0.943\pm0.002$, respectively. In contrast, no $\log$-singularity is present in the complementary channel $M_3^c$, whereas a significantly weaker $\log$-singularity of $ 0.096\pm0.002$ is observed for $(M_3\otimes A_{1/2})^c$. This pronounced asymmetry between a channel and its complementary channel directly accounts for the observed positivity of the coherent information. Further experimental details on the process tomography, eigenvalue spectra, and entropic properties are provided in the SM~\cite{SM}.

\begin{figure*}[htbp]
      \centering
    \begin{minipage}[t]{1\textwidth}
        \centering
          \hspace{-0.05\textwidth}
        \begin{overpic}[width=\linewidth]{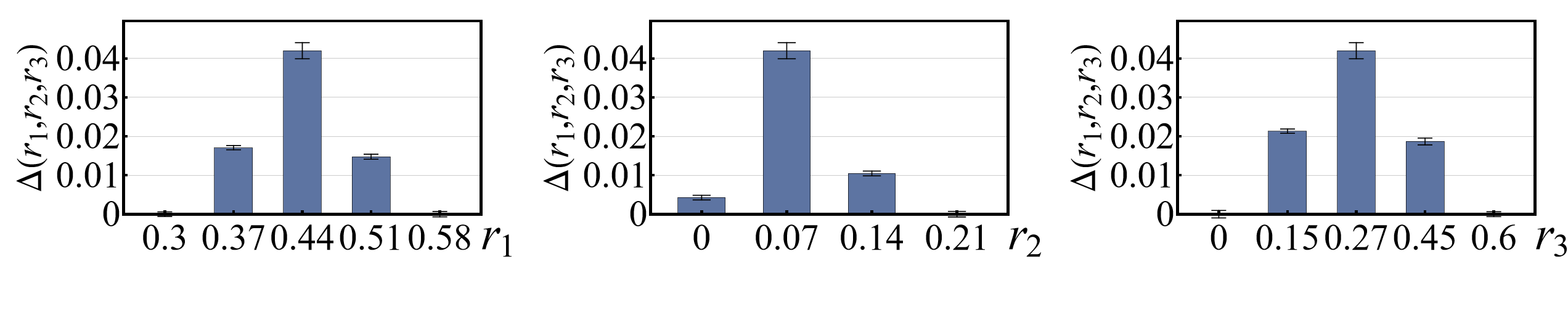}
            \put(18,1){\bfseries (a)} 
            \put(52,1){\bfseries (b)}
            \put(85,1){\bfseries (c)}
        \end{overpic}
        \label{fig:witnessresult-a}
    \end{minipage}

    \caption{\justifying\emph{Experimental observation of coherent information nonadditivity for the joint channel $M_3\otimes A_{1/2}$.} For the input state $\rho(r_1, r_2, r_3)$, we vary one parameter while fixing the other two at their respective optimal values and experimentally measure the corresponding value of $\Delta(r_1, r_2, r_3)$ (blue bars). The error bars are obtained via Monte Carlo simulations.}
    \label{fig:nonadditivity}
\end{figure*}

The $\log$-singularity properties of the channel $M_3\otimes A_{1/2}$ give rise to nonadditive quantum communication behavior~\cite{siddhu2021entropic}. We experimentally demonstrate this feature by directly measuring the coherent information for both the joint channel and the individual channels. Specifically, the coherent information of $M_3\otimes A_{1/2}$ is obtained by performing quantum state tomography on the output state corresponding to the input states of the form $\rho(r_1, r_2, r_3)=r_1|00\>\<00|+r_2|01\>\<01|+(1-r_1-r_2)(\sqrt{1-r_3}|20\>+\sqrt{r_3}|11\>)(\sqrt{1-r_3}\<20|+\sqrt{r_3}\<11|)$, with tunable parameters $r_i$, $i=1,2,3$. In contrast, the coherent information of $M_3$ is measured by preparing its optimal single-channel input state $\sigma=(1-u)|0\rangle\langle0|+u|2\rangle\langle2|$, with $u=0.445$. The average fidelity of the prepared input states is $0.993\pm 0.001$, demonstrating high-precision control of entangled states in a six-dimensional Hilbert space. Since the theoretical coherent information of the amplitude damping channel $A_{1/2}$ vanishes, we independently characterize its experimental realization by quantum process tomography and numerically optimize it over all the input states to determine its coherent information (see SM~\cite{SM} for details). As a result, we obtain $Q^{(1)}(M_3)=0.655\pm0.003$ and $Q^{(1)}(A_{1/2})=0$, which are consistent with the theoretical expectations.

To quantify nonadditivity, we introduce $\Delta(r_1,r_2,r_3)=Q^{(1)}(M_3\otimes A_{1/2})-Q^{(1)}(M_3)-Q^{(1)}(A_{1/2})$, which characterizes the deviation from the strong additivity of coherent information. Theoretically, the coherent information of $M_3\otimes A_{1/2}$ is maximized at $r_1=0.44$, $r_2=0.07$, and $r_3=0.27$. In Figs.~\ref{fig:nonadditivity}(a–c), we plot $\Delta(r_1,r_2,r_3)$ as a function of each parameter while the remaining two are fixed at their optimal values. A positive $\Delta(r_1,r_2,r_3)$ is observed over a broad parameter region, namely, $r_1\in[0.37,0.51]$, $r_2\in[0,0.14]$, and $r_3\in[0.14,0.46]$, providing clear experimental evidence of nonadditive quantum communication. The maximal value $\Delta(r_1,r_2,r_3)_{\max}=0.042\pm0.002$ is obtained near the optimal parameters, corresponding to a statistically significant violation of strong additivity. These results provide compelling experimental evidence that entropic singularities can induce coherent-information nonadditivity.

\medskip
\textit{Discussion.---} We have experimentally demonstrated quantum communication beyond additive limits through the joint use of a qutrit platypus channel and a qubit amplitude damping channel. By combining the precise control of high-dimensional photonic entanglement with engineered quantum channels, we directly resolved the entropy responses of both the channels and their complementary environments. Our results reveal a pronounced asymmetry in the logarithmic entropy response, where the output entropy develops substantially stronger singular behavior than the corresponding environmental entropy does. This entropic imbalance gives rise to positive coherent information and further enables clear superadditive enhancement under joint channel use.

More broadly, our work identifies entropic singularity as an experimentally accessible physical mechanism underlying quantum communication activation. The ability to reconstruct and manipulate singular entropy landscapes provides a new framework for exploring collective effects in quantum communication and suggests promising directions toward more general quantum activation phenomena.

\bigskip
\begin{acknowledgements}
This work was supported by the NSFC (No.~U25D8007, No.~62322513, No.~12374338, No.~12574402, and No.~12350006), the Quantum Science and Technology-National Science and Technology Major Project (No.~2021ZD0301200), the Fundamental Research Funds for the Central Universities (WK2030250138), Anhui Provincial Natural Science Foundation (No.~2408085JX002), Anhui Province Science and Technology Innovation Project (No.~202423r06050004), Guizhou Provincial Major Scientific and Technological Program XKBF (2025)010, 009, and Xiaomi Young Scholar. This work was partially carried out at the USTC Center for Micro and Nanoscale Research and Fabrication. 
\end{acknowledgements}

\bibliography{ref}

@article{smith2008quantum,
  title={Quantum communication with zero-capacity channels},
  author={Smith, Graeme and Yard, Jon},
  journal={Science},
  volume={321},
  number={5897},
  pages={1812--1815},
  year={2008},
  publisher={American Association for the Advancement of Science}
}

@article{hastings2009superadditivity,
  title={Superadditivity of communication capacity using entangled inputs},
  author={Hastings, Matthew B},
  journal={Nature Physics},
  volume={5},
  number={4},
  pages={255--257},
  year={2009},
  publisher={Nature Publishing Group UK London}
}

@article{smith2009extensive,
  title={Extensive nonadditivity of privacy},
  author={Smith, Graeme and Smolin, John A},
  journal={Physical Review Letters},
  volume={103},
  number={12},
  pages={120503},
  year={2009},
  publisher={APS}
}

@article{leditzky2018dephrasure,
  title={Dephrasure channel and superadditivity of coherent information},
  author={Leditzky, Felix and Leung, Debbie and Smith, Graeme},
  journal={Physical review letters},
  volume={121},
  number={16},
  pages={160501},
  year={2018},
  publisher={APS}
}

@article{leditzky2023generic,
  title={Generic nonadditivity of quantum capacity in simple channels},
  author={Leditzky, Felix and Leung, Debbie and Siddhu, Vikesh and Smith, Graeme and Smolin, John A},
  journal={Physical review letters},
  volume={130},
  number={20},
  pages={200801},
  year={2023},
  publisher={APS}
}

@article{holevo1973bounds,
  title={Bounds for the quantity of information transmitted by a quantum communication channel},
  author={Holevo, Alexander Semenovich},
  journal={Problemy Peredachi Informatsii},
  volume={9},
  number={3},
  pages={3--11},
  year={1973},
  publisher={Russian Academy of Sciences, Branch of Informatics, Computer Equipment and~…}
}

@article{devetak2005private,
  title={The private classical capacity and quantum capacity of a quantum channel},
  author={Devetak, Igor},
  journal={IEEE Transactions on Information Theory},
  volume={51},
  number={1},
  pages={44--55},
  year={2005},
  publisher={IEEE}
}

@article{shannon1948mathematical,
  title={A mathematical theory of communication},
  author={Shannon, Claude E},
  journal={The Bell system technical journal},
  volume={27},
  number={3},
  pages={379--423},
  year={1948},
  publisher={Nokia Bell Labs}
}

@article{schumacher1997sending,
  title={Sending classical information via noisy quantum channels},
  author={Schumacher, Benjamin and Westmoreland, Michael D},
  journal={Physical Review A},
  volume={56},
  number={1},
  pages={131},
  year={1997},
  publisher={APS}
}

@article{holevo2002capacity,
  title={The capacity of the quantum channel with general signal states},
  author={Holevo, Alexander S},
  journal={IEEE Transactions on Information Theory},
  volume={44},
  number={1},
  pages={269--273},
  year={2002},
  publisher={IEEE}
}

@article{bennett1996mixed,
  title={Mixed-state entanglement and quantum error correction},
  author={Bennett, Charles H and DiVincenzo, David P and Smolin, John A and Wootters, William K},
  journal={Physical Review A},
  volume={54},
  number={5},
  pages={3824},
  year={1996},
  publisher={APS}
}

@article{bennett1999entanglement,
  title={Entanglement-assisted classical capacity of noisy quantum channels},
  author={Bennett, Charles H and Shor, Peter W and Smolin, John A and Thapliyal, Ashish V},
  journal={Physical Review Letters},
  volume={83},
  number={15},
  pages={3081},
  year={1999},
  publisher={APS}
}

@article{cai2004quantum,
  title={Quantum privacy and quantum wiretap channels},
  author={Cai, Ning and Winter, Andreas and Yeung, Raymond W},
  journal={problems of information transmission},
  volume={40},
  number={4},
  pages={318--336},
  year={2004},
  publisher={Springer}
}

@article{lloyd1997capacity,
  title={Capacity of the noisy quantum channel},
  author={Lloyd, Seth},
  journal={Physical Review A},
  volume={55},
  number={3},
  pages={1613},
  year={1997},
  publisher={APS}
}

@article{barnum2002quantum,
  title={On quantum fidelities and channel capacities},
  author={Barnum, Howard and Knill, Emanuel and Nielsen, Michael A.},
  journal={IEEE Transactions on Information Theory},
  volume={46},
  number={4},
  pages={1317--1329},
  year={2002},
  publisher={IEEE}
}

@article{divincenzo1998quantum,
  title={Quantum-channel capacity of very noisy channels},
  author={DiVincenzo, David P and Shor, Peter W and Smolin, John A},
  journal={Physical Review A},
  volume={57},
  number={2},
  pages={830},
  year={1998},
  publisher={APS}
}

@article{macchiavello2016detecting,
  title={Detecting lower bounds to quantum channel capacities},
  author={Macchiavello, Chiara and Sacchi, Massimiliano F},
  journal={Physical Review Letters},
  volume={116},
  number={14},
  pages={140501},
  year={2016},
  publisher={APS}
}

@article{schumacher1996quantum,
  title={Quantum data processing and error correction},
  author={Schumacher, Benjamin and Nielsen, Michael A},
  journal={Physical Review A},
  volume={54},
  number={4},
  pages={2629},
  year={1996},
  publisher={APS}
}

@footnote{SM,
	note = {See Supplemental Material for more details.},
}

@article{siddhu2021entropic,
  title={Entropic singularities give rise to quantum transmission},
  author={Siddhu, Vikesh},
  journal={Nature Communications},
  volume={12},
  number={1},
  pages={5750},
  year={2021},
  publisher={Nature Publishing Group UK London}
}

@article{shor9604006quantum,
  title={Quantum error-correcting codes need not completely reveal the error syndrome,(1996)},
  author={Shor, Peter W and Smolin, John A},
  journal={arXiv preprint quantph/9604006}
}

@article{smith2007degenerate,
  title={Degenerate quantum codes for Pauli channels},
  author={Smith, Graeme and Smolin, John A},
  journal={Physical review letters},
  volume={98},
  number={3},
  pages={030501},
  year={2007},
  publisher={APS}
}

@article{smith2009can,
  title={Can nonprivate channels transmit quantum information?},
  author={Smith, Graeme and Smolin, John A},
  journal={Physical Review Letters},
  volume={102},
  number={1},
  pages={010501},
  year={2009},
  publisher={APS}
}

@article{elkouss2015superadditivity,
  title={Superadditivity of private information for any number of uses of the channel},
  author={Elkouss, David and Strelchuk, Sergii},
  journal={Physical Review Letters},
  volume={115},
  number={4},
  pages={040501},
  year={2015},
  publisher={APS}
}

@article{oppenheim2008quantum,
  title={For quantum information, two wrongs can make a right},
  author={Oppenheim, Jonathan},
  journal={Science},
  volume={321},
  number={5897},
  pages={1783--1784},
  year={2008},
  publisher={American Association for the Advancement of Science}
}

@article{smith2011quantum,
  title={Quantum communication with Gaussian channels of zero quantum capacity},
  author={Smith, Graeme and Smolin, John A and Yard, Jon},
  journal={Nature Photonics},
  volume={5},
  number={10},
  pages={624--627},
  year={2011},
  publisher={Nature Publishing Group UK London}
}

@article{brandao2012does,
  title={When does noise increase the quantum capacity?},
  author={Brand{\~a}o, Fernando GSL and Oppenheim, Jonathan and Strelchuk, Sergii},
  journal={Physical review letters},
  volume={108},
  number={4},
  pages={040501},
  year={2012},
  publisher={APS}
}

@article{lim2019activation,
  title={Activation and superactivation of single-mode Gaussian quantum channels},
  author={Lim, Youngrong and Takagi, Ryuji and Adesso, Gerardo and Lee, Soojoon},
  journal={Physical Review A},
  volume={99},
  number={3},
  pages={032337},
  year={2019},
  publisher={APS}
}

@article{horodecki2005secure,
  title={Secure key from bound entanglement},
  author={Horodecki, Karol and Horodecki, Micha{\l} and Horodecki, Pawe{\l} and Oppenheim, Jonathan},
  journal={Physical review letters},
  volume={94},
  number={16},
  pages={160502},
  year={2005},
  publisher={APS}
}

@article{leung2014maximal,
  title={Maximal privacy without coherence},
  author={Leung, Debbie and Li, Ke and Smith, Graeme and Smolin, John A},
  journal={Physical review letters},
  volume={113},
  number={3},
  pages={030502},
  year={2014},
  publisher={APS}
}

@article{leditzky2023platypus,
  title={The platypus of the quantum channel zoo},
  author={Leditzky, Felix and Leung, Debbie and Siddhu, Vikesh and Smith, Graeme and Smolin, John A},
  journal={IEEE Transactions on Information Theory},
  volume={69},
  number={6},
  pages={3825--3849},
  year={2023},
  publisher={IEEE}
}

@article{yu2020experimental,
  title={Experimental observation of coherent-information superadditivity in a dephrasure channel},
  author={Yu, Shang and Meng, Yu and Patel, Raj B and Wang, Yi-Tao and Ke, Zhi-Jin and Liu, Wei and Li, Zhi-Peng and Yang, Yuan-Ze and Zhang, Wen-Hao and Tang, Jian-Shun and others},
  journal={Physical Review Letters},
  volume={125},
  number={6},
  pages={060502},
  year={2020},
  publisher={APS}
}

\end{document}